\documentclass[twocolumn,journal]{IEEEtran}
\usepackage[T1]{fontenc}
\usepackage{color}
\usepackage{array}
\usepackage{float}
\usepackage{multirow}
\usepackage{amsmath}
\usepackage{amsthm}
\usepackage{amssymb}
\usepackage{stackrel}
\usepackage{graphicx}
\usepackage[unicode=true,
 bookmarks=true,bookmarksnumbered=true,bookmarksopen=true,bookmarksopenlevel=1,
 breaklinks=false,pdfborder={0 0 0},pdfborderstyle={},backref=false,colorlinks=false]
 {hyperref}
\hypersetup{pdftitle={Your Title},
 pdfauthor={Your Name},
 pdfpagelayout=OneColumn, pdfnewwindow=true, pdfstartview=XYZ, plainpages=false}

\makeatletter

\providecommand{\tabularnewline}{\\}
\floatstyle{ruled}
\newfloat{algorithm}{tbp}{loa}
\providecommand{\algorithmname}{Algorithm}
\floatname{algorithm}{\protect\algorithmname}

\theoremstyle{plain}
\newtheorem{thm}{\protect\theoremname}
\theoremstyle{plain}
\newtheorem{lem}[thm]{\protect\lemmaname}
\theoremstyle{remark}
\newtheorem{rem}[thm]{\protect\remarkname}

\usepackage[caption=false,font=footnotesize]{subfig}
\usepackage{graphicx}
\usepackage{cite}

\@ifundefined{showcaptionsetup}{}{%
 \PassOptionsToPackage{caption=false}{subfig}}
\usepackage{subfig}
\makeatother

\providecommand{\lemmaname}{Lemma}
\providecommand{\remarkname}{Remark}
\providecommand{\theoremname}{Theorem}

\begin{document}
\title{Learning Sparse Graphs via Majorization-Minimization for Smooth Node
Signals}
\author{Ghania Fatima, Aakash Arora, Prabhu Babu and Petre Stoica}
\maketitle
\begin{abstract}
In this letter, we propose an algorithm for learning a sparse weighted
graph by estimating its adjacency matrix under the assumption that
the observed signals vary smoothly over the nodes of the graph. The
proposed algorithm is based on the principle of majorization-minimization
(MM), wherein we first obtain a tight surrogate function for the graph
learning objective and then solve the resultant surrogate problem
which has a simple closed form solution. The proposed algorithm does
not require tuning of any hyperparameter and it has the desirable
feature of eliminating the inactive variables in the course of the
iterations - which can help speeding up the algorithm. The numerical
simulations conducted using both synthetic and real world (brain-network)
data show that the proposed algorithm converges faster, in terms of
the average number of iterations, than several existing methods in
the literature.
\end{abstract}


\section{Introduction and literature review}

Graph learning has recently gained widespread attention in the fields
of signal processing and machine learning. For example, graph learning
techniques can be used for the analysis of brain-networks \cite{key-1,key-2},
social networks \cite{key-4}, urban traffic networks \cite{key-5},
and can also be used to perform unsupervised learning \cite{key-6},
clustering \cite{key-7,key-8,key-9} etc. The data in the aforementioned
applications can be represented as signals that reside on the graph,
wherein each signal is assigned to a node of the graph, and the pairwise
relationships between the nodes are encoded as edge weights \cite{key-10,key-11}.
Learning the underlying graph structure helps in visualizing and describing
the interrelationships in the data. However, learning graphs from
large dimensional data sets is numerically challenging, so there is
a need for developing scalable and computationally efficient graph
learning algorithms.

An undirected weighted graph is often represented as $\small\mathcal{G}\left(\mathcal{V},\mathcal{E},\mathbf{W}\right)$,
where $\small\mathcal{V}=\left[1,2,\ldots,p\right]$ denotes the set
of nodes in the graph, $\mathcal{\small E\subseteq}\mathcal{V}\times\mathcal{V}$
is the set of edges, and $\small\mathbf{W}\in\mathbb{R}_{+}^{p\times p}$
is the adjacency matrix of the graph, the $\left(i,j\right)^{\mathrm{th}}$
entry of which takes a positive value if the edge corresponding to
the $i^{\mathrm{th}}$ and the $j^{\mathrm{th}}$ nodes is an element
of $\mathcal{E}$ or zero if the corresponding edge does not belong
to $\mathcal{E}$. The data matrix (which is the collection of $n$
different realizations of the graph signals) is denoted by $\small\mathbf{X}=[\mathbf{x}_{1},\mathbf{x}_{2},\mathbf{x}_{3},\ldots,\mathbf{x}_{p}]^{^{T}}\in\mathbb{R}^{p\times n}$,
where $\mathbf{x}_{p}\in\mathbb{R}^{n}$ contains the $n$ realizations
of the signal observed at node $p$. Given $\small\mathbf{X}$, learning
the graph is the problem of finding the edge weights from the data,
i.e., recovering the adjacency matrix $\small\mathbf{W}$ from $\small\mathbf{X}$
\cite{key-13}. To facilitate graph learning, two most common assumptions
made in the literature are that the underlying signals vary smoothly
(i.e., the difference between the signals at adjacent nodes is small)
and that the graph to be learned is sparse (i.e., $\small\mathbf{W}$
is a sparse matrix)\cite{key-13,key-14}. Under these assumptions,
different formulations of the graph learning problem have been proposed
\cite{key-2,key-13,key-14,key-15,key-16,key-17}. The graph learning
algorithm in \cite{key-14} learns the Laplacian matrix $\small\mathbf{L}$
(defined as $\small\mathbf{L}\triangleq\mathrm{diag}(\mathbf{W}\mathbf{1})-\mathbf{W}$,
where the vector $\small\mathbf{W}\mathbf{1}$ is usually referred
as the node degree vector) using a factor analysis model in which
smoothness is enforced by imposing a Gaussian probabilistic prior
on a set of latent variables that represent the graph signals. In
\cite{key-15}, the proposed method estimates the graph Laplacian
under structural constraints, which can also be interpreted as the
maximum a posteriori estimator of a Gaussian-Markov random field model.
The paper \cite{key-13} proposed the following convex method to directly
learn $\mathbf{W}$ (without invoking the Laplacian) and reported
that it has better performance than other existing methods (see \cite{key-13}
for more details on the formulation and notation):
\begin{equation}
\small\begin{aligned}\underset{\mathbf{W}\in\mathbb{R}_{+}^{p\times p}}{\mathrm{min}}\; & \left\Vert \mathbf{W}\circ\mathbf{D}\right\Vert _{1,1}-\alpha\boldsymbol{1}^{T}\mathrm{log}\left(\mathbf{W}\mathbf{1}\right)+\frac{\beta}{2}\left\Vert \mathbf{W}\right\Vert _{F}^{2}\\
\mathrm{s.t.}\; & \mathbf{W}=\mathbf{W}^{T},\:\mathrm{diag}(\mathbf{W})=0,
\end{aligned}
\label{eq:1}
\end{equation}
where $\small\mathbf{D}\in\mathbb{R}_{+}^{p\times p}$ is the pairwise
distance matrix whose elements are defined as $\small D_{i,j}=\left\Vert \mathbf{x}_{i}-\mathbf{x}_{j}\right\Vert _{2}^{2}$.
The first-term in the objective of (\ref{eq:1}) measures the smoothness
of the graph, the second-term (which is a log-barrier on the node
degree vector) ensures that the node degrees are positive and prevents
the edges associated with a node from all becoming zero (thus preventing
the learned graph from having isolated nodes) and the Frobenius norm
term $\small\frac{\beta}{2}\left\Vert \mathbf{W}\right\Vert _{F}^{2}$
controls the sparsity of the graph. The problem in (\ref{eq:1}) is
convex and has been extensively studied in the literature on graph
learning. Several algorithms have been developed to solve (\ref{eq:1}),
including the primal-dual (PD) algorithm in \cite{key-13}, the proximal
gradient (PG) algorithm in \cite{key-18}, the fast dual proximal
gradient (FDPG) algorithm in \cite{key-19}, and the linearized alternating
direction method of multipliers (ADMM) algorithm in \cite{key-20}.
All the aforementioned algorithms are global minimizers of (\ref{eq:1}),
however they differ in their convergence speed and their requirements
for selecting one or more search hyperparameters. For instance, the
PG method requires the selection of the step size, and the ADMM requires
the choice of Lagrangian multipliers. The PD algorithm was observed
to be relatively slow, whereas the FDPG (which is an accelerated version
of PG) was reported in the literature to be one of the fastest available
algorithms; however, the FDPG also requires the choice of some hyperparameters
and failing to choose the optimal value of those parameters can slow
down the method.

In this letter, we propose a new algorithm based on the MM approach
to solve (\ref{eq:1}). Unlike most of the state-of-the-art methods,
our algorithm does not require tuning any hyperparameter and in our
experience it enjoys faster convergence than the other methods. Moreover,
the proposed algorithm has the distinct feature of eliminating the
inactive elements of the weight matrix in the course of the iterations,
i.e. once a weight is found to be zero at an iteration it remains
zero throughout the future iterations of the algorithm. As a result,
the inactive elements in $\mathbf{W}$ can be removed and a smaller
dimensional optimization problem can be solved at the subsequent iterations.
This particular feature should be handy for high-dimensional sparse
graphs.

Section II of the letter introduces the details of the proposed algorithm,
Section III presents the numerical simulation results, and Section
IV concludes the paper.

\section{Proposed MM algorithm}

In this section, we first briefly describe the MM technique, then
provide a detailed derivation of the proposed algorithm, followed
by a brief discussion on its computational complexity and convergence.

\subsection{MM principle}

The MM technique consists of two steps: a majorization step and a
minimization step. In the majorization step, a surrogate function
$g\left(\mathbf{x}|\mathbf{x}^{\left(k\right)}\right)$ is constructed
such that it tightly upperbounds the objective function $\small f\left(\mathbf{x}\right)$
(to be minimized over $\mathbf{x}$) at a point $\mathbf{x}^{\left(k\right)}$
(the index $k$ denotes the iteration number):
\begin{equation}
\small g\left(\mathbf{x}|\mathbf{x}^{\left(k\right)}\right)\geq f\left(\mathbf{x}\right)\;\mathrm{and\;}\small g\left(\mathbf{x}^{\left(k\right)}|\mathbf{x}^{\left(k\right)}\right)=f\left(\mathbf{x}^{\left(k\right)}\right).\label{eq:2-1}
\end{equation}
In the minimization step, the surrogate function $\small g\left(\mathbf{x}|\mathbf{x}^{\left(k\right)}\right)$
is minimized to obtain the next update for $\mathbf{x}$:
\begin{equation}
\small\mathbf{x}^{\left(k+1\right)}=\mathrm{arg}\:\underset{\mathbf{x}}{\mathrm{min}}\;g\left(\mathbf{x}|\mathbf{x}^{\left(k\right)}\right).\label{eq:4-1}
\end{equation}
From (\ref{eq:2-1}) and (\ref{eq:4-1}), it can be easily deduced
that:
\begin{equation}
\small f\left(\mathbf{x}^{\left(k+1\right)}\right)\leq g\left(\mathbf{x}^{\left(k+1\right)}|\mathbf{x}^{\left(k\right)}\right)\leq g\left(\mathbf{x}^{\left(k\right)}|\mathbf{x}^{\left(k\right)}\right)=f\left(\mathbf{x}^{\left(k\right)}\right),\label{eq:5-1}
\end{equation}
which shows that the sequence $\small\left\{ f\left(\mathbf{x}^{\left(k\right)}\right)\right\} $
generated via MM is non-increasing. See \cite{key-25,key-21} for
more details on the MM technique and its applications in signal processing.

\subsection{Proposed algorithm}

Since $\small\mathbf{W}$ is a symmetric matrix and its diagonal elements
are zero, we can reformulate the objective of (\ref{eq:1}) in a vectorized
form as a function of a weight vector $\mathbf{w}$ consisting of
the upper triangular elements of $\small\mathbf{W}$:
\begin{equation}
\small\begin{aligned}\underset{\mathbf{w}\in\mathbb{R}_{+}^{m}}{\mathrm{min}}\; & \left\{ f(\mathbf{w})\triangleq2\mathbf{w}^{T}\mathbf{d}-\alpha\boldsymbol{1}^{T}\mathrm{log}\left(\mathbf{S}\mathbf{w}\right)+\beta\left\Vert \mathbf{w}\right\Vert _{2}^{2}\right\} ,\end{aligned}
\label{eq:2}
\end{equation}
where $\small\mathbf{d}\in\mathbb{R}_{+}^{m}$ denotes a vector made
of the elements lying above the main diagonal of the matrix $\small\mathbf{D}$,
$m=p(p-1)/2$, and $\small\mathbf{S}=\left[\mathbf{s}_{1},\mathbf{s}_{2},\ldots,\mathbf{s}_{p}\right]^{T}\in\mathbb{R}_{+}^{p\times m}$
is a binary matrix such that $\small\mathbf{S}\mathbf{w}=\mathbf{W}\mathbf{1}$.
Problem (\ref{eq:2}) can be rewritten as:
\begin{equation}
\small\begin{aligned}\underset{\mathbf{w}\in\mathbb{R}_{+}^{m}}{\mathrm{min}}\; & 2\mathbf{w}^{T}\mathbf{d}-\alpha\stackrel[i=1]{p}{\sum}\mathrm{log}\left(\mathbf{s}_{i}^{T}\mathbf{w}\right)+\beta\left\Vert \mathbf{w}\right\Vert _{2}^{2}.\end{aligned}
\label{eq:7-2}
\end{equation}
In the following, we devise an MM algorithm to find the global minimizer
of (\ref{eq:7-2}). In the first step we need to construct a tight
surrogate function for the objective in (\ref{eq:7-2}). To this end,
we prove the following lemma.
\begin{lem}
For vectors $\mathbf{s}_{i}$ and $\mathbf{w}$ with positive elements,
the convex function $\small-\mathrm{log}\left(\mathbf{s}_{i}^{T}\mathbf{w}\right)$
in the variable $\mathbf{w}$ can be upperbounded at a given point
$\mathbf{w}^{\left(k\right)}$ as 
\begin{equation}
\small-\mathrm{log}(\mathbf{s}_{i}^{T}\mathbf{w})\leq-\stackrel[j=1]{m}{\sum}\frac{s_{i,j}w_{j}^{\left(k\right)}}{\mathbf{s}_{i}^{T}\mathbf{w}^{\left(k\right)}}\mathrm{log}\left(\frac{\mathbf{s}_{i}^{T}\mathbf{w}^{\left(k\right)}}{w_{j}^{\left(k\right)}}w_{j}\right)\label{eq:8-1}
\end{equation}
with equality for $\mathbf{w}=\mathbf{w}^{\left(k\right)}$ (hereafter
$s_{i,j}$ denotes the $j^{th}$ element of $\mathbf{s}_{i}$).
\end{lem}
\begin{IEEEproof}
\noindent We first consider the function $\small-\mathrm{log}\left(x\right)$
(for $x>0$), which is convex in $x$. From Jensen's inequality we
have:
\begin{equation}
\small-\mathrm{log}\left(\stackrel[j=1]{m}{\sum}\alpha_{j}x_{j}\right)\leq-\stackrel[j=1]{m}{\sum}\alpha_{j}\mathrm{log}\left(x_{j}\right),\label{eq:9-1}
\end{equation}
where $\left\{ \alpha_{j}\right\} $ are scalars satisfying $\small\alpha_{j}\geq0$
and $\small\stackrel[j=1]{m}{\sum}\alpha_{j}=1$. Since the elements
of $\mathbf{s}_{i}$ and $\mathbf{w}$ are positive, we can choose
$\small\alpha_{j}\triangleq\frac{s_{i,j}w_{j}^{\left(k\right)}}{\mathbf{s}_{i}^{T}\mathbf{w}^{\left(k\right)}}$
and $\small x_{j}\triangleq\frac{\mathbf{s}_{i}^{T}\mathbf{w}^{\left(k\right)}}{w_{j}^{\left(k\right)}}w_{j}$
in (\ref{eq:9-1}) to get the desired inequality:
\begin{equation}
\small\begin{aligned}-\mathrm{log}(\mathbf{s}_{i}^{T}\mathbf{w})\leq-\stackrel[j=1]{m}{\sum}\frac{s_{i,j}w_{j}^{\left(k\right)}}{\mathbf{s}_{i}^{T}\mathbf{w}^{\left(k\right)}}\mathrm{log}\left(\frac{\mathbf{s}_{i}^{T}\mathbf{w}^{\left(k\right)}}{w_{j}^{\left(k\right)}}w_{j}\right).\end{aligned}
\end{equation}
\end{IEEEproof}
\noindent Using Lemma 1, we obtain the following majorizing function
for $\small f(\mathbf{w})$: 
\begin{equation}
\small\begin{aligned}f(\mathbf{w})\leq2\mathbf{w}^{T}\mathbf{d}-\alpha\stackrel[i=1]{p}{\sum}\stackrel[j=1]{m}{\sum}\frac{s_{i,j}w_{j}^{\left(k\right)}}{\mathbf{s}_{i}^{T}\mathbf{w}^{\left(k\right)}}\mathrm{log}\left(\frac{\mathbf{s}_{i}^{T}\mathbf{w}^{\left(k\right)}}{w_{j}^{\left(k\right)}}w_{j}\right)\\
+\beta\left\Vert \mathbf{w}\right\Vert _{2}^{2}\triangleq g\left(\mathbf{w}|\mathbf{w}^{\left(k\right)}\right).
\end{aligned}
\label{eq:12}
\end{equation}
Thus, the surrogate minimization problem is:
\begin{equation}
\small\begin{aligned}\underset{\mathbf{w}\in\mathbb{R}_{+}^{m}}{\mathrm{min}}\; & g\left(\mathbf{w}|\mathbf{w}^{\left(k\right)}\right)\end{aligned}
\label{eq:7-1}
\end{equation}
which we solve iteratively to obtain the minimizer of (\ref{eq:7-2}).
In Fig. 1, we illustrate the tightness of the surrogate function for
a simple example with a three-dimensional weight vector. The graph
signal for this example is generated as $\small\mathbf{x}_{p}\sim\mathcal{N}\left(\mathbf{0},\mathbf{L}^{\dagger}+\sigma^{2}\mathbf{I}_{n}\right)$,
where $\small\mathbf{L}^{\dagger}$ is the pseudo-inverse of the Laplacian
$\small\mathbf{L}$ of the ground-truth graph and $\sigma$ is $0.1$.
In Fig. 1(a), we plot $\small f(\mathbf{w})$ and $\small g\left(\mathbf{w}|\mathbf{w}^{\left(k\right)}\right)$
for $\small\mathbf{w}^{\left(k\right)}=\left[0.2\:0.2\:0.2\right]^{T}$
over $w_{1}$ and $w_{2}$ (with $w_{3}$ fixed at $0.2$). In Fig.
1(b), we show the one-dimensional cross-section of Fig. 1(a) where
$w_{2}$ is also fixed at $0.2$. It can be seen from Fig. 1(a) and
1(b) that the surrogate function tightly upperbounds $f\left(\mathbf{w}\right)$.

\noindent Using the expression in (\ref{eq:12}) for $\small g\left(\mathbf{w}|\mathbf{w}^{\left(k\right)}\right)$,
the surrogate optimization problem can be written as:
\begin{equation}
\small\begin{aligned}\underset{\mathbf{w}\in\mathbb{R}_{+}^{m}}{\mathrm{min}}\;2\mathbf{w}^{T}\mathbf{d}-\alpha\stackrel[i=1]{p}{\sum}\stackrel[j=1]{m}{\sum}\frac{s_{i,j}w_{j}^{\left(k\right)}}{\mathbf{s}_{i}^{T}\mathbf{w}^{\left(k\right)}}\mathrm{log}\left(\frac{\mathbf{s}_{i}^{T}\mathbf{w}^{\left(k\right)}}{w_{j}^{\left(k\right)}}w_{j}\right)+\beta\left\Vert \mathbf{w}\right\Vert _{2}^{2}.\end{aligned}
\label{eq:5}
\end{equation}

\begin{center}
\begin{figure}
\begin{centering}
\subfloat[]{\begin{centering}
\includegraphics[width=4.2cm,height=3.4cm]{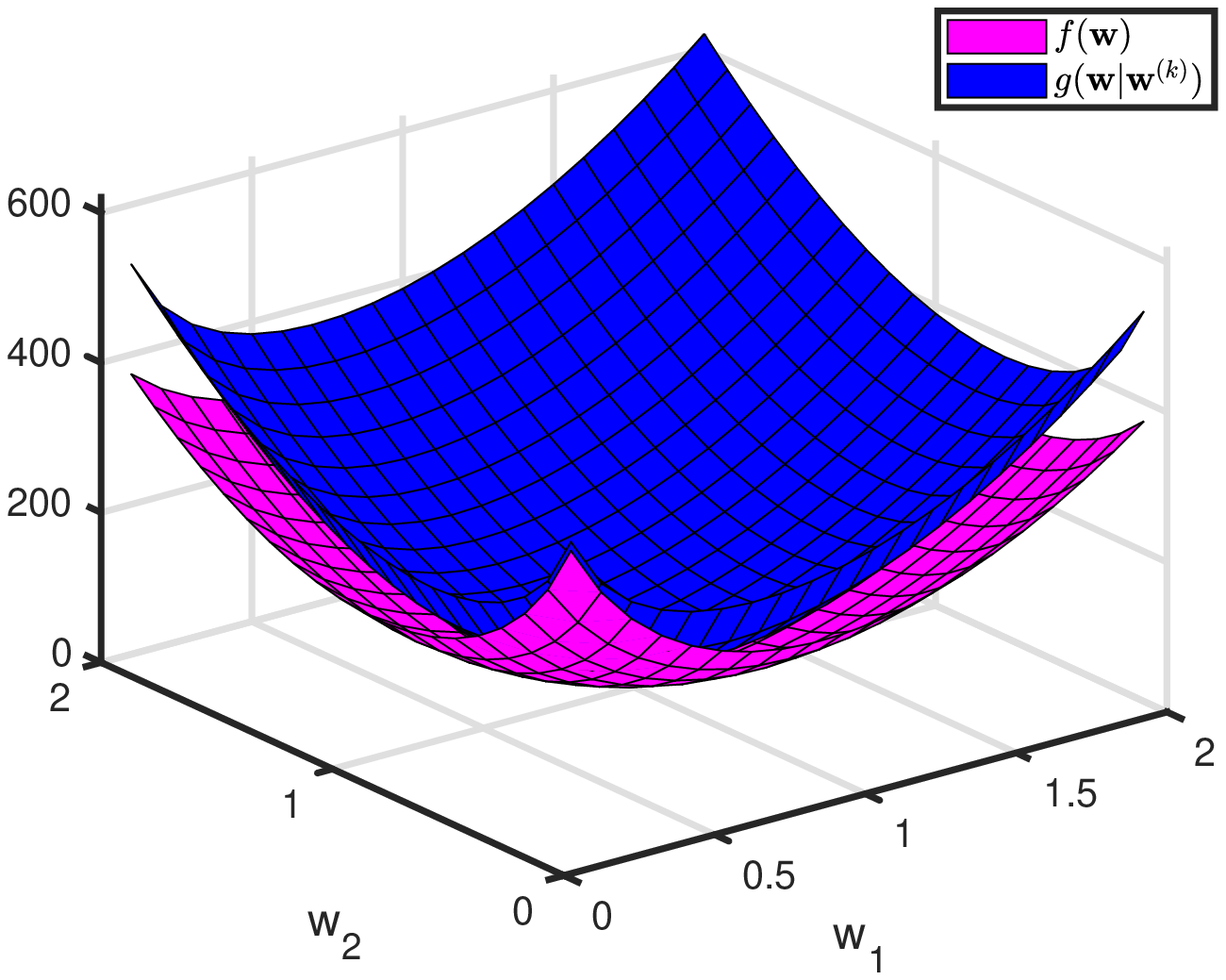}
\par\end{centering}
}\subfloat[]{\begin{centering}
\includegraphics[width=4.2cm,height=3.4cm]{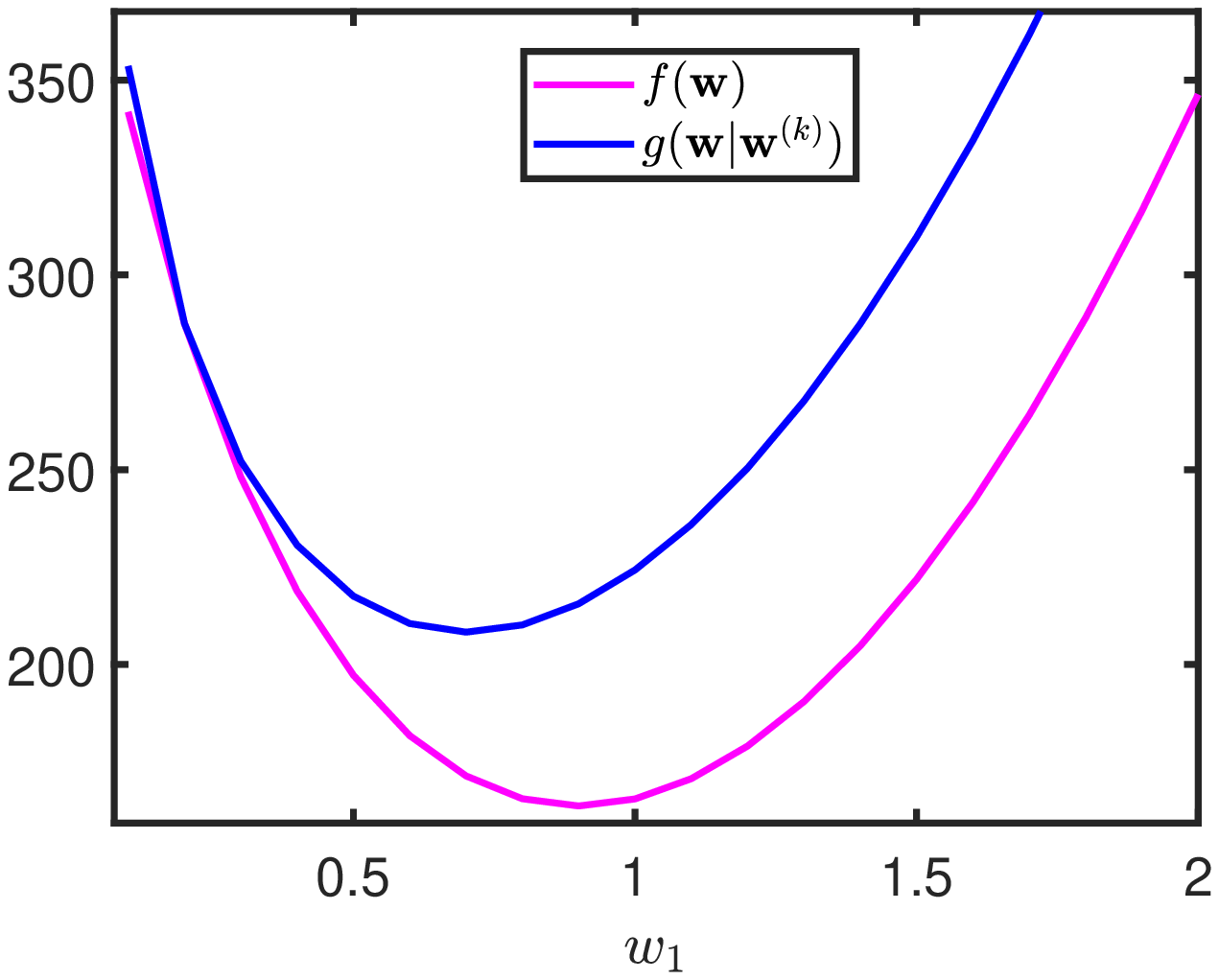}
\par\end{centering}
}
\par\end{centering}
\caption{The objective function $f(\mathbf{w})$ and the surrogate function
$g\left(\mathbf{w}|\mathbf{w}^{\left(k\right)}\right)$ at $\mathbf{w}^{\left(k\right)}=\left[0.2\:0.2\:0.2\right]^{T}$.
(a) $f(w_{1},w_{2}|w_{3}=0.2)$ and $g(w_{1},w_{2}|w_{3}=0.2|\mathbf{w}^{\left(k\right)})$
vs $w_{1}$ and $w_{2}$. (b) $f(w_{1}|w_{2}=0.2,w_{3}=0.2)$ and
$g(w_{1}|w_{2}=0.2,w_{3}=0.2|\mathbf{w}^{\left(k\right)})$ vs $w_{1}$.}
\end{figure}
\par\end{center}

\noindent The problem in (\ref{eq:5}) is separable in the elements
$w_{j}$'s of $\mathbf{w}$ and can be solved in a parallel manner:

\begin{equation}
\begin{aligned}\small\begin{aligned}\underset{w_{j}\in\mathbb{R}_{+}}{\mathrm{min}}\; & 2w_{j}d_{j}-\alpha\stackrel[i=1]{p}{\sum}\frac{s_{i,j}w_{j}^{\left(k\right)}}{\mathbf{s}_{i}^{T}\mathbf{w}^{\left(k\right)}}\mathrm{log}\left(\frac{\mathbf{s}_{i}^{T}\mathbf{w}^{\left(k\right)}}{w_{j}^{\left(k\right)}}w_{j}\right)+\beta w_{j}^{2}\end{aligned}
\\
(j=1,\ldots,n).
\end{aligned}
\label{eq:6}
\end{equation}
The first-order derivative of the objective in (\ref{eq:6}) is given
by:
\begin{equation}
\small2d_{j}+2\beta w_{j}-\alpha\stackrel[i=1]{p}{\sum}\frac{s_{i,j}w_{j}^{\left(k\right)}}{\mathbf{s}_{i}^{T}\mathbf{w}^{\left(k\right)}}\frac{1}{w_{j}}=0.\label{eq:7}
\end{equation}
Using the notation $\small c_{j}\triangleq\alpha\stackrel[i=1]{p}{\sum}\frac{s_{i,j}w_{j}^{\left(k\right)}}{\mathbf{s}_{i}^{T}\mathbf{w}^{\left(k\right)}}$,
(\ref{eq:7}) can be written as:
\begin{equation}
\small2d_{j}w_{j}+2\beta w_{j}^{2}-c_{j}=0.\label{eq:8}
\end{equation}
Solving (\ref{eq:8}), we get the following update for $w_{j}$ (which
is the positive root of the quadratic equation in (\ref{eq:8})):
\begin{equation}
\small w_{j}^{\left(k+1\right)}=\frac{-2d_{j}+\sqrt{4d_{j}^{2}+8\beta c_{j}}}{4\beta}\label{eq:9}
\end{equation}

\begin{rem}
From the above expressions for $c_{j}$ and $w_{j}^{\left(k+1\right)}$,
it can be inferred that if $c_{j}$ is zero, then the update in (\ref{eq:9})
will be zero as well and, moreover, it will remain zero in any subsequent
iteration. This is an important characteristic of our proposed algorithm,
in which we initialize $\mathbf{w}^{\left(0\right)}$ as an all-one
vector and then eliminate variables which become inactive, thereby
reducing the problem dimensionality in subsequent iterations and consequently
reducing the computational complexity.

The pseudo-code of the proposed algorithm (the vanilla version without
variable elimination) is given in Algorithm 1.
\end{rem}
\begin{algorithm}
\caption{Pseudo-code for MM-based graph learning algorithm}

\textbf{Input: $\mathbf{X}$, $\epsilon=10^{-4}$}

\textbf{Initialize:} $\small\mathbf{w}^{\left(0\right)}=\mathbf{1}$

\textbf{Iterate}: Given $\small\mathbf{w}^{\left(k\right)}$, do the
$\small\left(k+1\right)^{\mathrm{th}}$ step:

\hspace{20pt}Calculate $\small\left\{ c_{j}\right\} _{j=1}^{m}$.

\hspace{20pt}Obtain $\small\mathbf{w}^{\left(k+1\right)}$ using
(\ref{eq:9}).

\textbf{Stop} If $\small\mathrm{abs}\left(\frac{f(\mathbf{w}^{\left(k\right)})-f(\mathbf{w}^{\left(k+1\right)})}{f(\mathbf{w}^{\left(k\right)})}\right)\leq\epsilon$.
\end{algorithm}

\subsection{Computational complexity and convergence of the MM algorithm}

The main computational step of the proposed algorithm is the calculation
of $\mathbf{S}\mathbf{w}$. The matrix $\mathbf{S}$ is sparse and
the calculation of $\mathbf{S}\mathbf{w}$ incurs a per iteration
cost of $O(p^{2})$, which is same as the per iteration cost of the
other state-of-the-art algorithms such as \cite{key-13,key-18,key-19,key-20}.

Since the proposed algorithm is based on MM principle, the sequence
$\small\left\{ f\left(\mathbf{w}^{\left(k\right)}\right)\right\} $
generated is non-increasing. The convergence of the sequence to a
stationary point of (\ref{eq:2}) can be established using the contents
of Section II-C of \cite{key-21} and the references therein.

\begin{figure*}
\begin{centering}
\subfloat[]{\begin{centering}
\includegraphics[width=5.75cm,height=4.25cm]{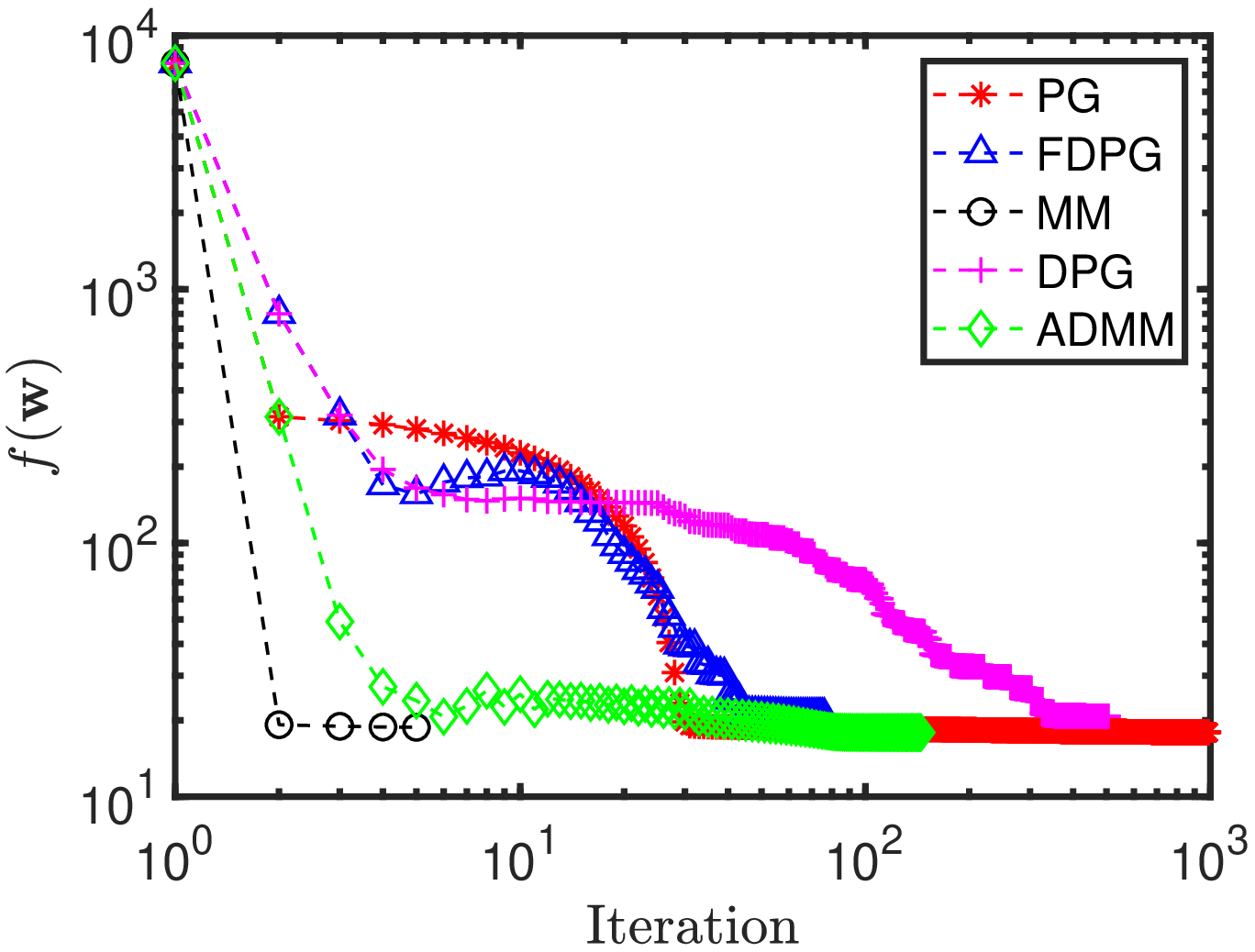}
\par\end{centering}

}\subfloat[]{\begin{centering}
\includegraphics[width=5.75cm,height=4.25cm]{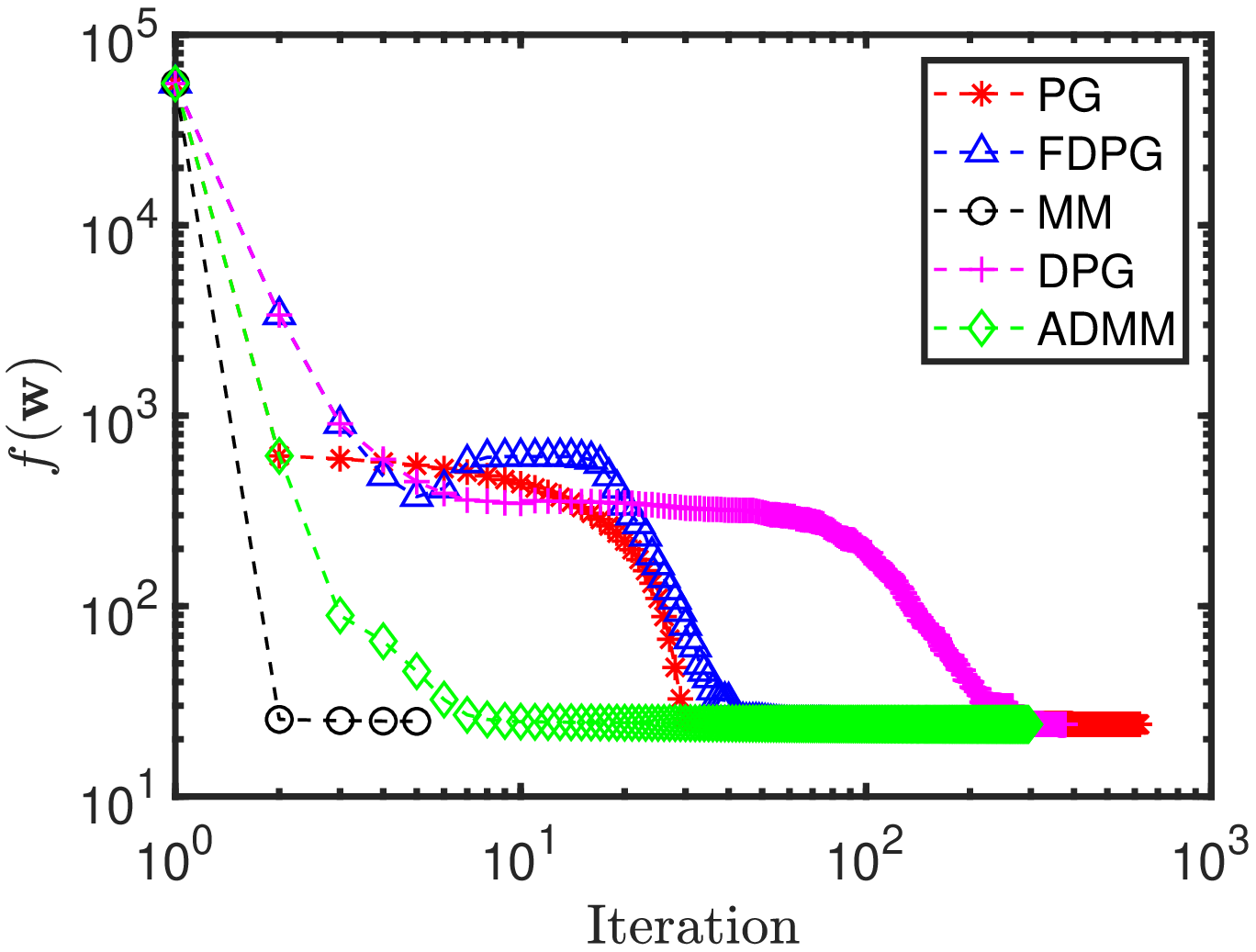}
\par\end{centering}
}\subfloat[]{\begin{centering}
\includegraphics[width=5.75cm,height=4.25cm]{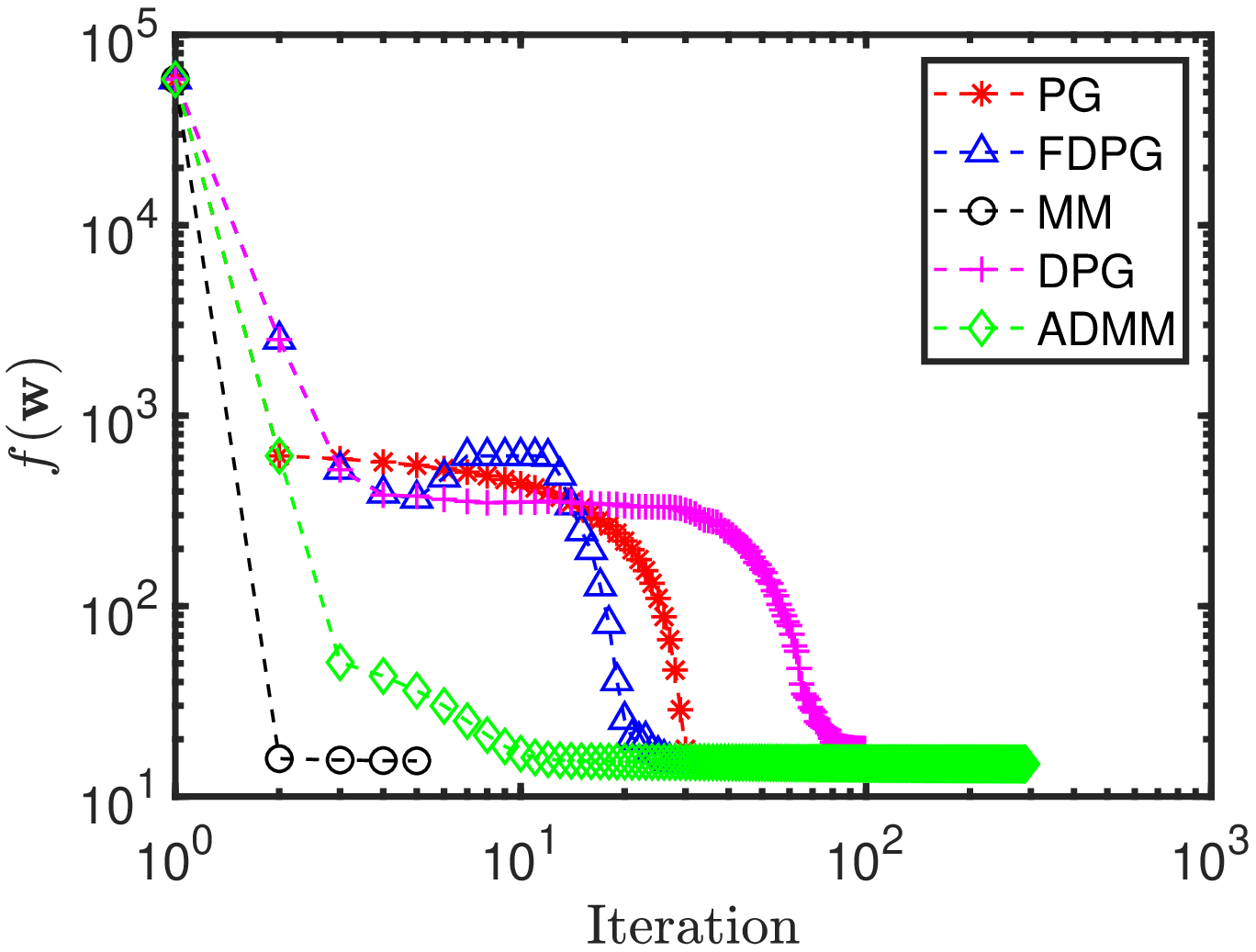}
\par\end{centering}
}\vfill{}
\subfloat[]{\begin{centering}
\includegraphics[width=5.75cm,height=4.25cm]{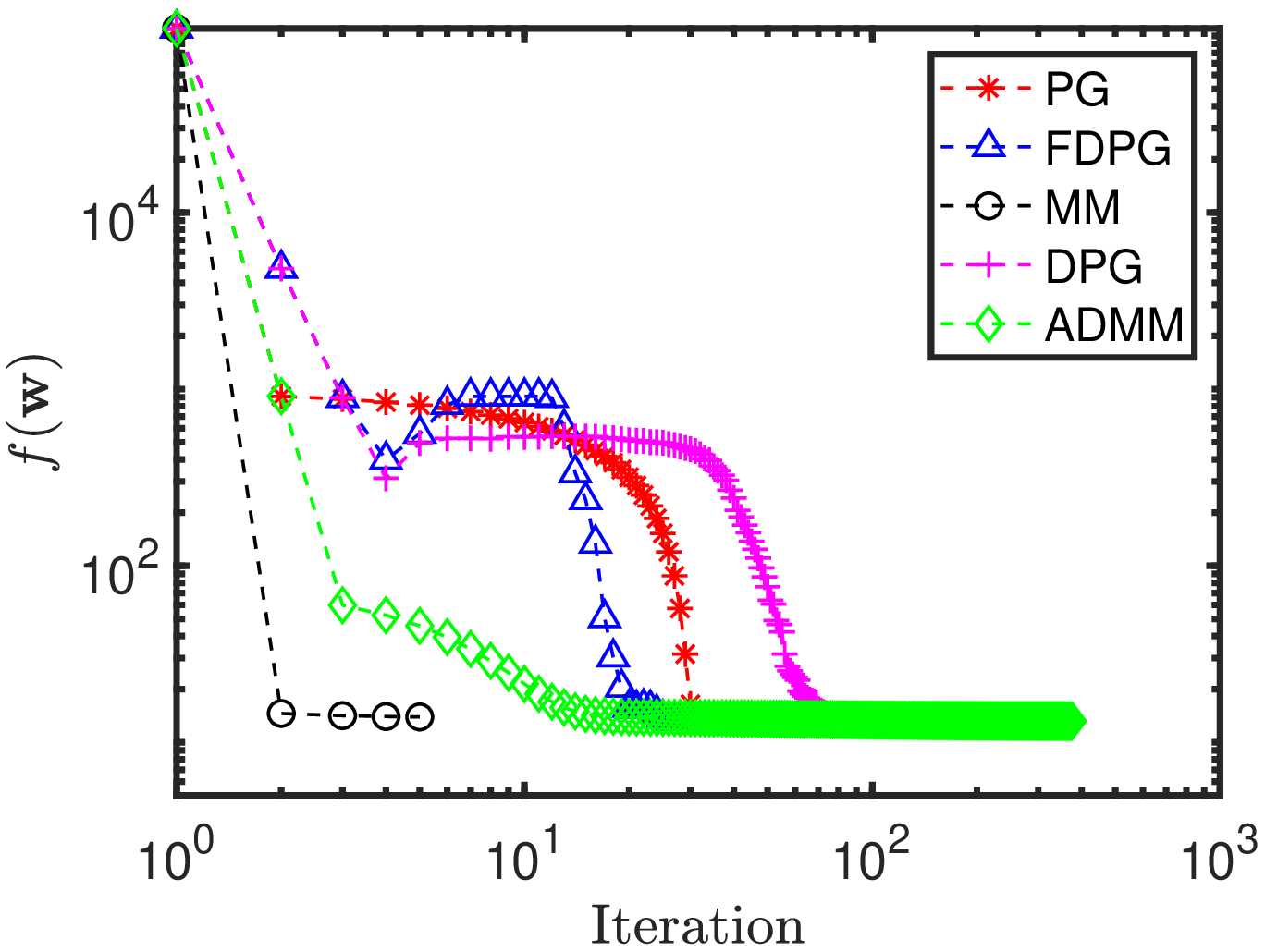}
\par\end{centering}
}\subfloat[]{\begin{centering}
\includegraphics[width=5.75cm,height=4.25cm]{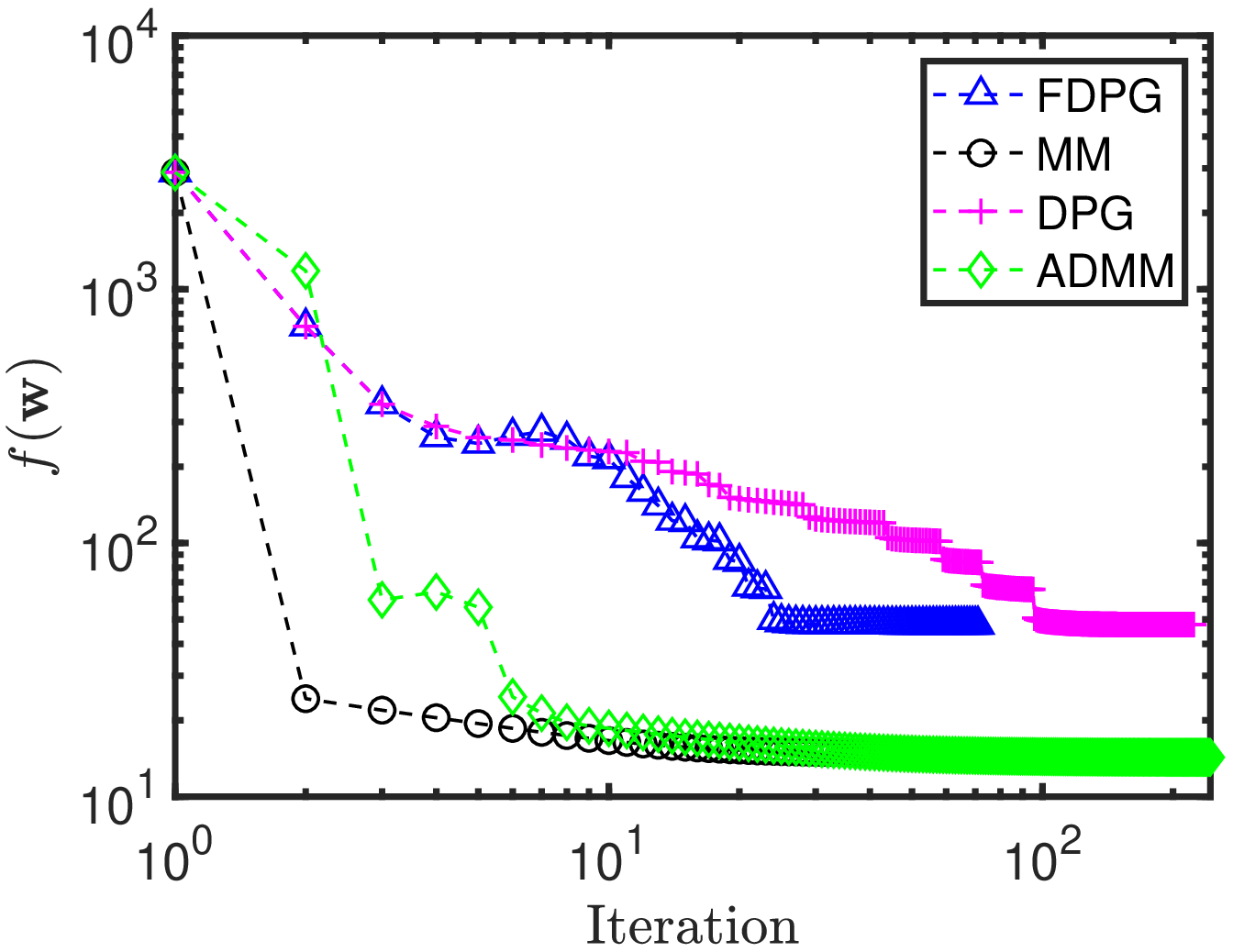}
\par\end{centering}
}\subfloat[]{\begin{centering}
\includegraphics[width=5.75cm,height=4.25cm]{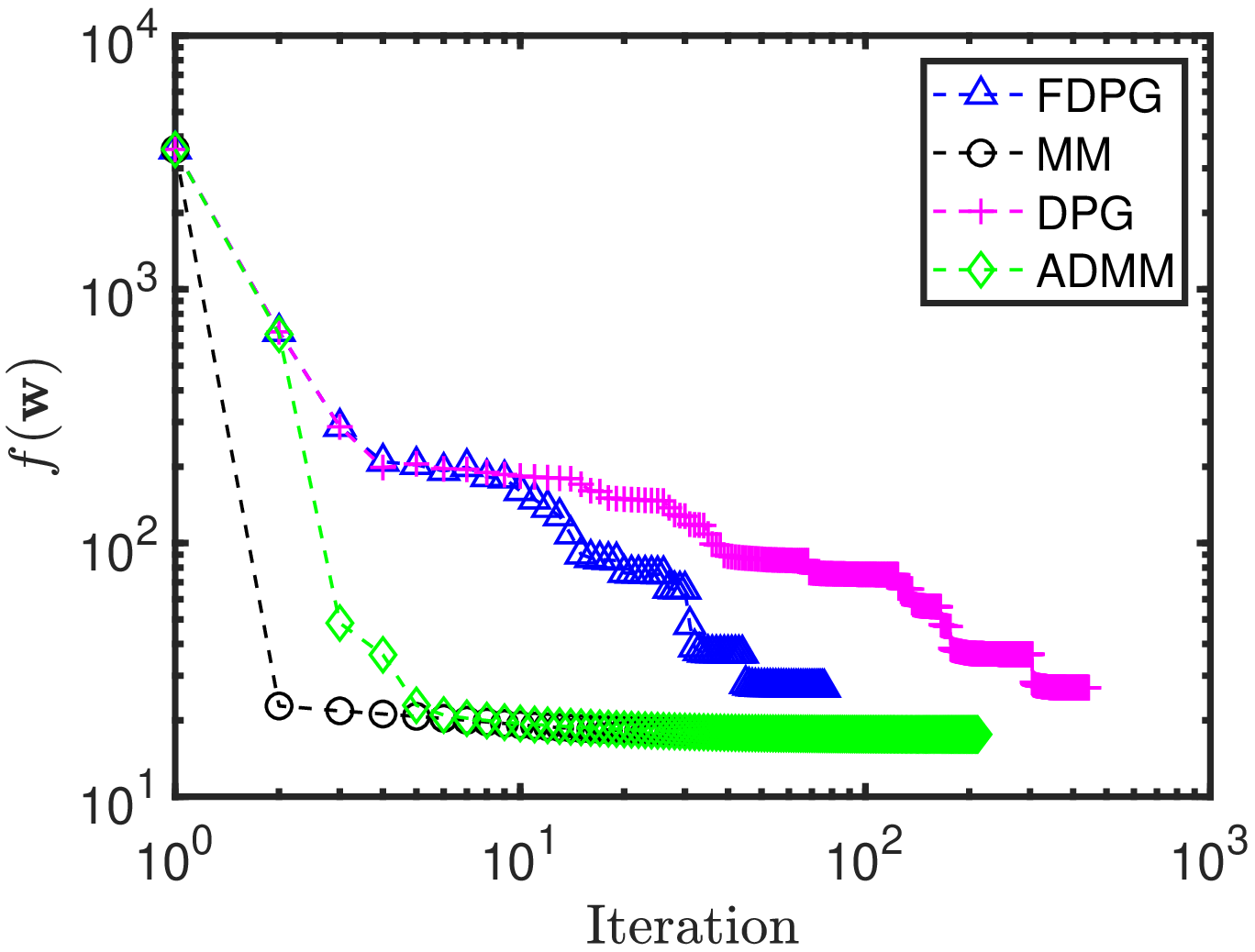}
\par\end{centering}
}
\par\end{centering}
\centering{}\caption{$f\left(\mathbf{w}\right)$ vs iteration number, for different graph
settings and for $n=1200$ samples of the graph signals. (a) ER graph
($100$ nodes). (b) ER graph ($200$ nodes). (c) SBM graph ($200$
nodes). (d) SBM graph ($300$ nodes). (e) Brain graph with $66$ ROIs
for Subject 3. (f) Brain graph with $66$ ROIs for Subject 5. Since
the numbers of iterations required by the PG algorithm (for real-world
graphs), and PD algorithm (for both synthetic and real-world graphs)
are significantly larger than for the others, we have omitted these
algorithms from the corresponding plots in this figure.}
\end{figure*}

\section{NUMERICAL SIMULATIONS}

In this section, the proposed algorithm is tested on both synthetic
graphs and a real-world graph (brain network \cite{key-24}). The
experiments are conducted using MATLAB on a PC with Intel i7 processor
and 16 GB RAM. The values of parameters $\alpha$ and $\beta$ are
pre-tuned for best graph recovery performance as in \cite{key-14}
and the smooth graph signals are also simulated as in \cite{key-14}.
All the simulations are done for the sample length of $n=1200$. The
performance of the proposed algorithm is compared with that of the
PD \cite{key-13}, PG \cite{key-18}, ADMM \cite{key-20}, and DPG
and FDPG \cite{key-19} algorithms\footnote{The following publicly available code has been used to implement these
algorithms: http://www.ece.rochester.edu/\~{ }gmateosb/code/FDPG.zip}. As suggested in \cite{key-20}, the penalty parameter and the step
size in case of ADMM are obtained via cross-validation by running
$1000$ iterations of the algorithm for a total of $1300$ possible
combinations of the parameters and choosing the combination giving
the least cross-validation error.

\subsection{Synthetic graph}

We consider two synthetic graph models: the Erdos-Renyi (ER) graph
model (with edge formation probability equal to $0.1$), and the 2-block
Stochastic Block Model (SBM) (with edge formation probability equal
to $0.3$ for nodes in the same block and $0.05$ for nodes in different
blocks), and generate ground-truth graphs corresponding to both models.
We plot the evolution of the objective function with the number of
iterations for each algorithm and for different numbers of nodes in
each model. Fig. 2(a) and 2(b) show the evolution of the objective
function with the number of iterations for the ER model with $p=100\;\mathrm{and}\;200$
nodes, whereas Fig. 2(c) and 2(d) correspond to the SBM model with
$p=200\;\mathrm{and}\;300$ nodes. From these figures, one can see
that the proposed algorithm has the fastest convergence rate in all
cases (also see Table I). The average computation time for the proposed
method is also shorter than that of the competing methods. For instance,
for the case of an ER graph (with $400$ nodes), the average computation
time is $0.1950$ sec for the proposed MM method, $0.2601$ sec for
FDPG, $0.9685$ sec for DPG, $1.1478$ sec for PG and $1.6196$ sec
for ADMM (the computation time for ADMM does not include the time
needed to choose the optimal search parameters).

\vspace{-5pt}

\subsection{Brain network}

As a real-world example, we consider the structural brain graph with
$p=66$ regions of interests (ROIs) (and the regional connection matrix
as given in \cite{key-24}) for six different subjects. The objective
function variation with the number of iterations for subject 3 and
subject 5 is shown in Fig. 2(e) and 2(f), respectively. Once again
the proposed algorithm needs the smallest number of iterations to
converge in comparison to the other algorithms. In Table I, we present
the average number of iterations needed by each algorithm, computed
from $100$ Monte Carlo simulations, for both real graphs (brain network
graphs for the remaining subjects 1, 2, 4 and 6) and several ER and
SBM synthetic graphs. It can be seen that the proposed method needs
the least number of iterations for both synthetic and real graphs.
\begin{center}
\begin{table}
\begin{centering}
\caption{Average number of iterations needed by each algorithm}
\par\end{centering}
\begin{center}
\resizebox{8.5cm}{2.5cm}{%
\centering{}%
\begin{tabular}{|>{\raggedright}m{2cm}|>{\centering}m{2.5cm}|>{\centering}m{0.75cm}|>{\centering}p{0.5cm}|>{\centering}p{0.6cm}|>{\centering}p{0.75cm}|>{\centering}p{0.5cm}|>{\centering}p{0.5cm}|}
\hline 
\multicolumn{1}{|>{\raggedright}m{2cm}}{} &  & \textbf{\footnotesize{}ADMM} & \textbf{\footnotesize{}PG} & \textbf{\footnotesize{}DPG} & \textbf{\footnotesize{}FDPG} & \textbf{\footnotesize{}PD} & \textbf{\footnotesize{}MM}\tabularnewline
\hline 
\hline 
\multirow{6}{2cm}{\textbf{\footnotesize{}SYNTHETIC GRAPH}} & \textbf{\footnotesize{}ER (100 nodes)} & {\footnotesize{}19} & {\footnotesize{}225} & {\footnotesize{}71} & {\footnotesize{}38} & {\footnotesize{}261} & \textbf{\textcolor{black}{\footnotesize{}6}}\tabularnewline
\cline{2-8} \cline{3-8} \cline{4-8} \cline{5-8} \cline{6-8} \cline{7-8} \cline{8-8} 
 & \textbf{\footnotesize{}ER (200 nodes)} & {\footnotesize{}11} & {\footnotesize{}32} & {\footnotesize{}180} & {\footnotesize{}46} & {\footnotesize{}500} & \textbf{\textcolor{black}{\footnotesize{}5}}\tabularnewline
\cline{2-8} \cline{3-8} \cline{4-8} \cline{5-8} \cline{6-8} \cline{7-8} \cline{8-8} 
 & \textbf{\footnotesize{}ER (300 nodes)} & {\footnotesize{}14} & {\footnotesize{}32} & {\footnotesize{}174} & {\footnotesize{}45} & {\footnotesize{}683} & \textbf{\textcolor{black}{\footnotesize{}5}}\tabularnewline
\cline{2-8} \cline{3-8} \cline{4-8} \cline{5-8} \cline{6-8} \cline{7-8} \cline{8-8} 
 & \textbf{\footnotesize{}ER (400 nodes)} & {\footnotesize{}17} & {\footnotesize{}33} & {\footnotesize{}146} & {\footnotesize{}39} & {\footnotesize{}845} & \textbf{\textcolor{black}{\footnotesize{}5}}\tabularnewline
\cline{2-8} \cline{3-8} \cline{4-8} \cline{5-8} \cline{6-8} \cline{7-8} \cline{8-8} 
 & \textbf{\footnotesize{}SBM (200 nodes)} & {\footnotesize{}14} & {\footnotesize{}32} & {\footnotesize{}88} & {\footnotesize{}28} & {\footnotesize{}902} & \textbf{\textcolor{black}{\footnotesize{}5}}\tabularnewline
\cline{2-8} \cline{3-8} \cline{4-8} \cline{5-8} \cline{6-8} \cline{7-8} \cline{8-8} 
 & \textbf{\footnotesize{}SBM (300 nodes)} & {\footnotesize{}19} & {\footnotesize{}33} & {\footnotesize{}75} & {\footnotesize{}26} & {\footnotesize{}1232} & \textbf{\textcolor{black}{\footnotesize{}5}}\tabularnewline
\hline 
\multirow{4}{2cm}{\textbf{\footnotesize{}REAL GRAPH}} & \textbf{\footnotesize{}Brain Graph (Subject 1)} & {\footnotesize{}122} & {\footnotesize{}1620} & {\footnotesize{}223} & {\footnotesize{}60} & {\footnotesize{}396} & \textbf{\textcolor{black}{\footnotesize{}53}}\tabularnewline
\cline{2-8} \cline{3-8} \cline{4-8} \cline{5-8} \cline{6-8} \cline{7-8} \cline{8-8} 
 & \textbf{\footnotesize{}Brain Graph (Subject 2)} & {\footnotesize{}120} & {\footnotesize{}5101} & {\footnotesize{}209} & {\footnotesize{}55} & {\footnotesize{}413} & \textbf{\textcolor{black}{\footnotesize{}53}}\tabularnewline
\cline{2-8} \cline{3-8} \cline{4-8} \cline{5-8} \cline{6-8} \cline{7-8} \cline{8-8} 
 & \textbf{\footnotesize{}Brain Graph (Subject 4)} & {\footnotesize{}128} & {\footnotesize{}3845} & {\footnotesize{}251} & {\footnotesize{}60} & {\footnotesize{}381} & \textbf{\textcolor{black}{\footnotesize{}52}}\tabularnewline
\cline{2-8} \cline{3-8} \cline{4-8} \cline{5-8} \cline{6-8} \cline{7-8} \cline{8-8} 
 & \textbf{\footnotesize{}Brain Graph (Subject 6)} & {\footnotesize{}108} & {\footnotesize{}1224} & {\footnotesize{}254} & {\footnotesize{}62} & {\footnotesize{}349} & \textbf{\textcolor{black}{\footnotesize{}51}}\tabularnewline
\hline 
\end{tabular}}
\end{center}
\end{table}
\par\end{center}

\section{CONCLUSION}

In this letter, we have developed a novel algorithm for graph learning
from smooth signals using the MM technique, wherein we construct a
simple surrogate function that tightly upperbounds the objective,
and which has a simple closed-form minimizer. An important feature
of our method is that once an edge weight becomes zero during the
course of the iteration, it remains zero at all subsequent iterations,
a property that can be used for variable elimination and therefore
the computation time reduction. The proposed algorithm, which is easy
to understand and code, decreases the objective function monotonically
and enjoys guaranteed convergence to the global optimum of the graph
learning objective without requiring the tuning of any search parameters.

\end{document}